\documentclass[prb,superscriptaddress,showpacs,amsmath,amssymb,twocolumn,floatfix]{revtex4}

\usepackage{graphicx}
\usepackage{dcolumn}
\usepackage{color}

\newcommand{\vc}[1]{{\mathbf #1}}

\begin{document}

\widetext

\title{ 
Geminal embedding scheme for optimal atomic basis set construction
in correlated calculations
}

\begin{abstract}
We introduce an efficient method to construct optimal 
and system adaptive
basis sets for
use in electronic structure and quantum Monte Carlo calculations. The
method is based on an 
embedding scheme in which a reference atom is singled out from its
environment, while the entire system (atom and environment) is
described by a Slater determinant or 
its antisymmetrized geminal power (AGP) extension. 
The embedding procedure described here allows for the systematic and 
consistent contraction of the primitive basis set into geminal
embedded orbitals (GEOs), with a dramatic reduction of the number of variational
parameters necessary to represent the many-body wave function, for a
chosen target accuracy.  
Within the variational Monte Carlo method, the Slater or AGP part is 
determined by a variational minimization of the energy of the  
whole system in presence of a flexible and accurate
Jastrow factor, representing most of the dynamical electronic
correlation. 
The resulting GEO basis set opens the way for a
fully controlled optimization of many-body wave functions in
electronic structure calculation of bulk 
materials, namely containing a large number of electrons and
atoms. We present applications on the water molecule, the volume
collapse transition in cerium, and the high-pressure liquid hydrogen.
\end{abstract}

\author{S. Sorella}
\email[]{sorella@sissa.it}
\affiliation{International School for Advanced Studies (SISSA) Via Beirut 2,4
  34014 Trieste, Italy and INFM Democritos National Simulation Center,
  Trieste, Italy} 
\author{N. Devaux}
\affiliation{Institut de Min\'eralogie, 
de Physique des Mat\'eriaux et de Cosmochimie,
Universit\'e Pierre et Marie Curie,
case 115, 4 place Jussieu, 75252, Paris cedex 05, France}
\author{M. Dagrada}
\email[]{mario.dagrada@impmc.upmc.fr}
\affiliation{Institut de Min\'eralogie, 
de Physique des Mat\'eriaux et de Cosmochimie,
Universit\'e Pierre et Marie Curie,
case 115, 4 place Jussieu, 75252, Paris cedex 05, France}
\author{G. Mazzola}
\email[]{gmazzola@phys.ethz.ch}
\affiliation{Theoretische Physik, ETH Zurich, 8093 Zurich, Switzerland}
\author{M. Casula}
\email[]{michele.casula@impmc.upmc.fr}
\affiliation{CNRS and Institut de Min\'eralogie,
de Physique des Mat\'eriaux et de Cosmochimie,
Universit\'e Pierre et Marie Curie,
case 115, 4 place Jussieu, 75252, Paris cedex 05, France}
\date{\today}






\maketitle

\section{Introduction}
\label{intro}

In \emph{ab initio} quantum chemistry and computational
condensed matter physics the optimization of the basis set has been,
just from the very beginning, a crucial 
ingredient for defining feasible algorithms that can provide
meaningful and converged physical and chemical properties in
electronic calculations. 

In density functional theory (DFT) calculations with periodic boundary
conditions, plane-wave (PW) basis sets are mostly used since their systematic
convergence can be controlled by just a single parameter, the PW
cutoff. This outweighs disadvantages such as the loss of a chemical
intuitive picture, and the need of pseudopotentials to smooth out the core region.
On the other hand, large-scale coarse-graining or $O(N)$ -$N$ being the 
total number of electrons- algorithms typically
require localized basis sets.\cite{cp2k}

In quantum chemistry calculations, mainly based on Gaussian type
orbitals (GTO), a tremendous effort has been done
to reduce the size of the localized basis set, while keeping the
same level of accuracy as the one of the corresponding primitive
basis. Indeed, the computational cost crucially depends on the
basis set size $L$, growing as fast as $L^4$, if the four-index
interaction integrals are fully evaluated. 
Very effective basis have been proposed, such as
Dunning's\cite{dunning, correlation_consistent2},
Peterson's\cite{peterson}, and Weigend's\cite{weigend}, which allow one to systematically
converge to the complete basis set (CBS) limit. Their construction
required a thorough analysis of the correlation effects in free atoms,
and their dependence on the orbital components, systematically added to the
basis set.

Efficient schemes to generate optimal atomic basis in a more
automatic way have been developed along the years. Those are
usually based on the diagonalization of the density matrix computed
for the atomic ground state, as firstly proposed by Alml\"of and
Taylor\cite{ano,ano2}. The resulting atomic natural orbitals (ANO) are
contractions of the Gaussian primitive basis set, provided automatically by the
density matrix diagonalization. To improve their transferability and
generate a more balanced basis set for molecular calculations, Widmark
and coworkers\cite{widmark1990,widmark1995,widmark2004} devised better
schemes, based on the 
diagonalization of a density matrix appropriately averaged over
several atomic states. An interesting recent development in the basis set
generation shows that a high-quality basis can be generated by
combining ANO orbitals obtained from the
density matrix of an atomic multiconfigurational self-consistent field
(MCSCF), with Gauss-Slater mixed primitive functions\cite{cyrus-gs}
optimized for the homonuclear dimers at the coupled cluster single
double (CCSD) level of theory. The resulting compact ANO-GS basis set is
particularly suited for quantum Monte Carlo (QMC) calculations\cite{ano-gs}.

The idea of using the ANOs to improve the convergence of the basis
set dates back to the seminal paper by
L\"owdin\cite{seminal_natural}. Weinhold and coworkes 
developed the ANO formalism to find a set of natural hybrid
orbitals (NHO)\cite{natural_hybrids} 
which are optimal not only for their convergence properties but also
because they allow a clearer interpretation of the chemical
bond\cite{natural_population_analysis} out of 
a quantum chemistry calculation in large basis sets, where the
chemical picture is not usually transparent. Since then, several
papers appeared, with the aim at finding the best scheme to generate
a minimal basis bearing all physical information on the local atom
embedded in a quantum
system\cite{Bader1981,cioslowsky_localized_natural_orbitals,cioslowsky_embedding,Mayer_minimal_basis_set,polarized_minimal_basis_set,intrinsic_minimal_atomic_basis,minimal_spillage_minimal_basis_set,knizia,local_basis_analysis}.   

In quantum Monte Carlo (QMC) calculations compact and efficient basis
sets are eagerly needed. In variational Monte Carlo (VMC), the task is to define  
a consistent many-body wave function,
namely a correlated ansatz providing the minimum possible energy
expectation value of the non-relativistic Hamiltonian with long-range
Coulomb forces and within the Born-Oppenheimer (BO) approximation.
Generally speaking, a correlated ansatz is made of a determinantal
part, usually the Slater determinant, which fulfills the antisymmetric
properties of electrons,  
multiplied by a Jastrow factor which takes into account the
dynamic correlation. While few basis functions, either
GTO\cite{casula_mol} or polynomials\cite{PhysRevB.70.235119}, are sufficient to define a good Jastrow factor,
the determinantal part remains the most complicated object to
develop, even if its atomic basis set expansion converges more rapidly
when the Kato cusp conditions\cite{kato} are fulfilled exactly by the Jastrow part.
Moreover, to have a good description of static correlations,
wave functions beyond the Jastrow-Slater form have been proposed, where
the Slater determinant is replaced by complete active space (CAS)
wave functions\cite{chemical_accuracy_QMC}, antisymmetrized geminal power (AGP) wave
functions\cite{casula_atom,casula_mol}, and pfaffians\cite{PhysRevLett.96.130201,PhysRevB.77.115112}.

It is well known that  a consistent QMC wave
function, with both the Jastrow and  
the determinantal parts simultaneously optimized within a given basis set,  
provides much better properties. Thanks to recent methodological
developments\cite{srh,umrigar,attaccalite_sorella}, the energy minimization of the determinantal
parameters can be systematically carried out. However,
in order to allow the simultaneous energy optimization of the Jastrow 
and the determinantal parts,  
it is extremely important to reduce the atomic basis
size $M_A$ of the determinant, because, as we will see in the following, the number
of variational parameters necessary to define  
a variational wave function scales as $M_A^2$. 

In this paper we introduce a method, based on a density
matrix embedding of the determinantal part, which allows
for a systematic reduction of 
the dimension of the atomic basis,  
by yielding atomic orbitals in an automatic and
almost black-box procedure. 
This method is an application of the concept of quantum
entanglement\cite{neumann} between 
a part of a system (A) interacting with the environment (B), where 
one represents A (B)  with a set of  $M_A$
 ($M_B$) orthogonal states labeled by 
the index $i$ ($j$) and writes down the wave function of the universe $U=A \cup B$ as:
\begin{equation} \label{psiu}
 |U \rangle = \sum\limits_{i=1,j=1}^{M_A,M_B} \psi_{ij} | i \otimes j \rangle.
\end{equation}
It is straightforward to show\cite{dmrg_review} that the optimal way to describe the universe 
by using only few $p << M_A$ states of the system {\em embedded in the universe} 
 is obtained by using the $p$  
eigenvectors corresponding to the  largest eigenvalues of the density matrix:
\begin{equation}\label{dmbar}
D_{ii^\prime} = \sum\limits_j \psi_{ij}^* \psi_{i^\prime j}.
\end{equation}
The approach above is essentially equivalent to the Schmidt decomposition of the 
rectangular matrix $\psi_{ij}$. It is extremely simple and general, and
it has been successfully applied in a variety of embedding
schemes, going from the celebrated density matrix renormalization
group (DMRG)\cite{dmrg} to the recent density matrix embedding
theory (DMET)\cite{chan,chan_quantum_chemistry}, and its simplified
density embedding
version\cite{density_embedding_theory,ab_initio_density_embedding}. 

Our method is based upon a modification of the previously 
described concept of embedding, where the universe is restricted to
the determinant of a geminal (or pairing) function, i.e. $|U \rangle =
\det\{\phi(\vc{r}_i,\vc{r}_j)\}$. In this approach,
the many-body coefficients $\psi_{ij}$ of Eq.~\ref{psiu} are replaced
by the $f_{ij}$ coefficients, which define the pairing function:
\begin{equation}
\label{pairing_embedding}
\phi(\vc{r},\vc{r}^\prime) = \sum_{i,j=1}^{M_A+M_B} f_{ij}\langle \vc{r} | i \rangle
\langle \vc{r}^\prime| j \rangle,
\end{equation}
where the $M_A + M_B$ states $|i\rangle$ are now a one-body atom-centered basis set, which  
spans the whole space ($A \cup B$), and is not restricted to be orthonormal.
Thus, in our approach the universe is represented by 
an AGP function, which includes the Slater determinant as the lowest
rank- i.e. $N/2$- limit of $f_{ij}$, when the $\phi(\vc{r},\vc{r}^\prime)$ becomes
equal to the standard one-body density matrix.
As we show in the paper, this formalism allows
one to define the embedding at the geminal level, by dealing with
2-body objects in a much simpler way than the direct 
N-body integration required to
generate the one-body density matrix in the general correlated case. 
The resulting geminal embedded orbitals (GEOs) represent an 
orthonormal basis set, and are defined as generalized hybrid orbitals, namely  
contractions of non-orthogonal atomic orbitals, optimally chosen to
minimize the basis set extension at a given target accuracy. We show
that for single determinant wave functions
the method presented here is superior to the ``standard'' natural
hybrid orbitals (NHOs)
generation (which leads to the so-called ``maximum-occupancy''
orbitals\cite{natural_hybrids}), thanks to
the efficiency of the present embedding scheme.

In practice, the GEOs can be generated from previous ``mean-field''
calculations, such as Hartree-Fock (HF) and DFT, or from previous
Jastrow correlated Slater determinant (JSD) or AGP (JAGP) quantum Monte Carlo calculations
carried out in an extended basis set. In the JSD and JAGP cases, the embedding
is performed for the determinantal part only. This procedure yields
GEOs which depend on the local environment, and are therefore
fundamentally different from the previously proposed ANOs basis sets,
as the latter are determined for free, isolated atoms. 
If the GEOs are taken at the single Slater determinant level, our
method is similar to the embedding proposed in Ref.~\onlinecite{link_orbitals_chan},
which is focused on fixing the best \emph{linking} orbitals from a target
region to a given environment. Here, the embedding is meant to give
the best orbitals for the \emph{target} region itself, where the
target is usually every single atomic site, but it is not necessarily limited to it.
Once determined at the DFT level, the GEOs coefficients can be further
optimized in a subsequent QMC 
energy minimization, efficiently performed in an optimally contracted
basis set and for a significantly reduced number of variational parameters.

The paper is organized as follows. In Sec.~\ref{method} we describe
the density matrix embedding and the way to find the optimal GEOs,
Sec.~\ref{results} shows how the scheme works in practice through
selected applications (the water molecule in Sec.~\ref{water},
the $\alpha$-to-$\gamma$ transition in solid
cerium in Sec.~\ref{cerium}, and the liquid hydrogen at high pressure in
Sec.~\ref{hydrogen}). Finally Sec.~\ref{conclusions} is devoted to the
concluding remarks.

\section{ Geminal embedded orbitals construction}
\label{method}

\subsection{Wave function form}
\label{wf}

We use the paramagnetic
Jastrow correlated Slater determinant (JSD) as ansatz in our solid state
calculations, with parameters determined to minimize the energy of the
scalar-relativistic first-principles Hamiltonian. Unless otherwise
specified, the full Coulomb
electron-ion interaction is replaced by a scalar-relativistic
Hartree-Fock energy-consistent pseudopotential of Burkatzki, Filippi, and Dolg (BFD)
type\cite{bfd,dolg_private}. 

The JSD wave function reads
\begin{equation}
\Psi_\textrm{JSD}({\bf R}_\textrm{el}) = \exp[-J({\bf R}_\textrm{el})]
\det[\psi_i^{MO}( \vc{r}_j^\uparrow)] \det[\psi_i^{MO}( \vc{r}_j^\downarrow)] , 
\label{JSD}
\end{equation}
where $1 \le i,j \le N/2$, and ${\bf R}_\textrm{el}=\{ {\bf r}_1^\uparrow,\ldots,
{\bf r}_{N/2}^\uparrow, {\bf r}_1^\downarrow,\ldots,
{\bf r}_{N/2}^\downarrow\}$  the many-body electron configuration,
with $N$ the total number of electrons.
$\psi_i^{MO}(\textbf{r})$ are orthonormal
molecular orbitals (MOs) each one   
occupied by opposite spin electrons. 
The orbitals
$\psi_i^{MO}(\textbf{r})$ are expanded in a GTO
basis set $\{\chi^\textrm{det}_j\}$, centered on the atomic nuclei, i.e.
\begin{equation}
\psi_i^{MO}(\textbf{r})=\sum_{j=1}^{M_A \times N_\textrm{atoms}} \mu_{ij}\chi^\textrm{det}_j (\textbf{r}),
\label{MO_expansion}
\end{equation}
where the sum in the above Equation runs over both the local basis set
($M_A$) and nuclear center indices ($N_\textrm{atoms}$).
The $\{\chi^\textrm{det}_{i}\}$ basis set is uncontracted (primitive)
with a system dependent size $M_A$.
The $\mu_{ij}$ and the exponents
of the primitive Gaussian basis set $\{\chi^\textrm{det}_j\}$ are
variational parameters. 
The primitive atomic basis $\chi^\textrm{det}_i(\vc{r})$ is not
constrained by any orthogonalization condition, namely the overlap matrix
$s_{ij} = \langle \chi^\textrm{det}_i | \chi^\textrm{det}_j \rangle$
is an arbitrary strictly positive definite matrix. 
Nevertheless the coefficients $\mu_{ij}$ can be  determined
in a way that the molecular
orbitals remain orthonormal: 
\begin{equation}
\langle \psi^{MO}_i | \psi^{MO}_j \rangle = \delta_{ij},
\label{orthonormal}
\end{equation}
namely, $\mu s \mu^\dag = I$. 
The first guess for $\psi_i^{MO}$
is provided by density functional theory (DFT) calculations in the
local density approximation (LDA),
performed in the same basis set.

To go beyond the JSD ansatz in Eq.(\ref{JSD}), we use its
JAGP extension in solid state calculations
wherever the Fermi level in the supercell is degenerate, and in molecular applications.
In the JAGP case, the wave function reads
\begin{equation}
\Psi_\textrm{JAGP}({\bf R}_\textrm{el}) = \exp[-J({\bf R}_\textrm{el})]
\det[\phi( \vc{r}_i^\uparrow,\vc{r}_j^\downarrow)]. 
\label{JAGP}
\end{equation}
The geminal function $\phi$ in Eq.(\ref{JAGP}) is written as:
\begin{equation}
\phi({\bf r},{\bf r}^\prime)= \sum_{i=1}^M \lambda^\textrm{AGP}_i \psi_i^{MO}({\bf
  r}) \psi_i^{MO}({\bf r}^\prime).
\label{pairing_MO}
\end{equation}
If $M=N/2$ and $\lambda^\textrm{AGP}_i=1$ for $i=1,\ldots,N/2$, the
expansion of Eq.(\ref{pairing_MO}) is equivalent to the 
single Slater determinant in Eq.(\ref{JSD}), which factorizes into up
and down components. However, to better describe static correlations
in molecular calculations,
$M$ can be larger to include orbitals above the HOMO level.
In the case of solid state calculations with degenerate shells at the
Fermi level, $M$ can be larger to comprise all degenerate
orbitals, with all the HOMO $\lambda^\textrm{AGP}_i$ taken equal and tiny.
One can prove that the AGP part of 
$\Psi_\textrm{JAGP}$ becomes then a linear combination of SDs, each
containing one degenerate orbital\cite{casula_atom}. In this way, the
shell degeneracy is correctly taken into account, and the symmetry of
the supercell is not broken. 

This variational ansatz has proven very accurate in a large
variety of \emph{ab initio} systems,  
molecules\cite{casula_mol,sorella2007,optimization_barborini_2012} and
solids\cite{sandro_silicon,mariapia_graphene,casula_fese}. 
The Jastrow factor $J$ is the one expanded over Gaussian basis set
orbitals, first introduced in Ref.~\onlinecite{casula_mol} and further
developed later (see e.g. \onlinecite{electrical_properties_Coccia_2012} and references therein). 
It is not detailed here, as it is not
the main focus of the present paper. Moreover, the embedding scheme
of the determinantal part devised here is very general, and can be
used in combination with other types of Jastrow
factors\cite{RPA_Jastrow,PhysRevB.53.9640,generic_Jastrow}, or directly on Slater determinants
generated by HF or DFT. 
Both DFT and quantum Monte Carlo
(QMC) calculations have been carried out using the 
$\textsc{TurboRVB}$ package\cite{turboRVB}.

\subsection{Geminal embedding scheme}
\label{embedding_scheme}

\subsubsection{General framework}
\label{general_framework}

The starting molecular orbitals $\psi^{MO}_i(\vc{r})$ are optimized in a finite 
localized basis, where each element $\chi^\textrm{det}_i(\vc{r})$ is
centered at a given 
atomic position $\vc{R}_i$ (see Eq.(\ref{MO_expansion})). 
The purpose of the present section is to determine a method that is able
to minimize the number of atomic basis elements for a fixed target
accuracy, once it is assumed that the original elements of the basis  
are given by localized orbitals.

Of course for large basis sets the location $\vc{R}_i$ of the 
atomic orbitals becomes 
an ill-defined concept because in principle a complete basis can be 
generated in any position of the space and not necessarily around a given 
atom. In the following we will consider a reasonable dimension for 
the original basis set $\{ \chi^\textrm{det}_i \}$ because: i) Large
localized basis have the well known problem to be highly redundant,
preventing a stable energy minimization of the many body wavefunction 
 within the Monte Carlo approach, which is the main application 
we will consider in the following; ii) Within this scheme it is usually enough
to consider rather small and well-conditioned basis sets 
(i.e. the corresponding overlap matrix has a condition number much
smaller than the inverse numerical relative precision)  to have quite
accurate physical results. 

We note that $\{ \chi^\textrm{det}_i \}$ is not restricted to be a
set of GTOs. Indeed, it can represent a more general atomic basis set,
over which the GEOs will be expanded. However, all the results
presented in this paper are obtained using GTOs as $\{
\chi^\textrm{det}_i \}$,
as already stated in Sec.~\ref{wf}.
In the following we do not restrict $\{ \chi^\textrm{det}_i \}$ to be
orthonormal, either.
We also take all functions described in this Section 
real, for the sake of simplicity, as it is not difficult to generalize
this derivation to the complex case.

We consider the general form of the geminal function in Eq.(\ref{pairing_MO}).
As we have seen, it can describe also a single Slater determinant,
with $\lambda^\textrm{AGP}_i=1$ for $i=1,\ldots,N/2$,
and $\{\lambda_i^\textrm{AGP}=0\}_{i > N/2}$. 
Like the Slater determinant, when the $\lambda^\textrm{AGP}_i$ 
are all equal, the geminal is  
invariant for all unitary transformations ${\mathcal O}$ of the 
molecular orbitals:
\begin{equation}
 \psi^{MO}_i \to \sum_j \mathcal{O}_{i,j} \psi^{MO}_j. 
\end{equation}
Therefore the geminal is a 
convenient representation of a Slater determinant, 
as it allows the use of  a metric
in the space of $ {\cal R}^3 \times {\cal R}^3$
pairing functions, much simpler than the many-body distance 
defined in the ${\cal R}^{3N}$ Hilbert space, and nevertheless without 
destroying the invariance upon unitary rotations $\mathcal{O}$. 
The same metric can be used in the more general form when
the SD becomes a true AGP (i.e. with a geminal of rank $> N/2$).

The embedding we propose here is done at the geminal level, by
left projecting Eq.(\ref{pairing_MO}) over a single atom centered at $\vc{R}$:
\begin{equation} \label{universe}
U_\textrm{proj}^\vc{R}({\bf r},{\bf r^\prime}) = \sum_i
\lambda^\textrm{AGP}_i  \psi^\textrm{proj}_{i,\vc{R}}( {\bf r}) \psi^{MO}_i (
{\bf r^\prime } ), 
\end{equation}
where $\psi^\textrm{proj}_{i,\vc{R}}( {\bf r})$ are 
obtained  by expanding the molecular orbitals $ \psi^{MO}_i (\bf r)$
on the atomic basis set $\{ \chi^\textrm{det}_i \}$
(Eq.(\ref{MO_expansion})) and considering 
in $\psi^\textrm{proj}_{i,\vc{R}}({\bf r})$ only
those components centered on the chosen atom, namely:
\begin{equation}
\psi^\textrm{proj}_{i,\vc{R}}( {\bf r})=\sum_{j | \vc{R}_j={\vc R}} \mu_{ij}\chi^\textrm{det}_j (\textbf{r}).
\label{geo_proj}
\end{equation}
Unless otherwise stated, hereafter we are going to omit the symbol $\vc{R}$ in
the projected quantities, for the sake of readability.
The left-projected geminal defined in Eq.(\ref{universe})
plays the role of the entangled
state corresponding to the system (selected atom) plus the 
environment (all atoms) in Eq.(\ref{psiu}), where now the indices labeling 
the system A and the environment B are replaced by positions 
${\bf r}$ and ${\bf r^\prime}$, respectively. 

We then determine the ``best'' geminal embedded atomic orbitals
$\phi^{GEO}_i ({\bf r})$ by  
representing the left-projected  geminal function $\sum_{i=1}^p
\phi^{GEO}_i ({\bf r}) \bar \psi_i({\bf r^\prime})$ in an optimally
reduced space, namely  
in terms of only $p<<M_A$ atomic natural hybrid orbitals centered on the
reference atom and of corresponding 
auxiliary molecular orbitals $\bar \psi_i ({\bf r^\prime})$ spanning all the system.
This can be achieved by a standard Schmidt decomposition, through a  
minimization of the Euclidean distance between the truncated and the projected 
geminal function.
 This minimization will be shown to be equivalent to diagonalize the 
density matrix kernel, defined, in analogy with Eq.(\ref{dmbar}), as:
\begin{equation}\label{truedm}
D_\textrm{proj}({\bf r},{\bf \bar r})=\int d{\bf r^\prime} U_\textrm{proj} ({\bf r}, {\bf r^\prime}) U_\textrm{proj}({\bf \bar r },{\bf r^\prime})
\end{equation}
The corresponding eigenvalues $w_i$  
may be related to the GEOs occupation and
their chemical reactivity. 

\subsubsection{Detailed procedure}
\label{detailed_procedure}

In the following we are going to explain our procedure to determine 
a substantial reduction of the basis dimension in more
details. We  rewrite the term in Eq.(\ref{universe}) by expanding it in the chosen 
 atomic basis: 
\begin{equation}
U_\textrm{proj}(\vc{r},\vc{r}^\prime) = \sum\limits_{k}^{M} \lambda^\textrm{AGP}_k 
\sum\limits_{i | \vc{R}_i=\vc{R}} \sum_j \mu_{ki} \mu_{kj} 
\chi^\textrm{det}_i(\vc{r}) \chi^\textrm{det}_j(\vc{r}^\prime)
\label{proj_det_mat} 
\end{equation} 
To shorthand the notation, Eq.(\ref{proj_det_mat}) can be also written in terms
of a matrix $\lambda$:
\begin{equation} 
U_\textrm{proj}(\vc{r},\vc{r}^\prime) = \sum\limits_{ij} \lambda_{ij} 
\chi^\textrm{det}_i(\vc{r}) \chi^\textrm{det}_j(\vc{r}^\prime),
\label{proj_det_mat_short}
\end{equation} 
where $\lambda_{ij}=[\mu^\dag \lambda^\textrm{AGP} \mu]_{ij}$\footnote{
Here the matrix $\lambda^\textrm{AGP}$ is a diagonal one with matrix 
elements $\lambda^\textrm{AGP}_k \delta_{k,k^\prime}$.}
 if the orbital
$\chi^\textrm{det}_i(\vc{r})$ is such that $\vc{R}_i = \vc{R}$,
while $\lambda_{ij}=0$ if $\vc{R}_i \ne \vc{R}$,
whereas the column index $j$ runs all over the atomic basis.

The term defined  in Eq.(\ref{proj_det_mat}) carries
information on the intra-atomic electronic structure affected by inter-atomic
interactions between the site $\vc{R}$ and its environment. The
inter-atomic interactions are explicitly kept by the left-partial
projection of the full density matrix. We found this embedding scheme
particularly effective to determine the best GEOs spanning an optimally
truncated Hilbert space.

We employ  the Schmidt decomposition  of
Eq.(\ref{proj_det_mat}) in a truncated space spanned by
$p$ terms only, as:
\begin{equation}
{\bar U_\textrm{proj}} (\vc{r},\vc{r}^\prime) = \sum \limits_{k=1}^p \phi_k^{GEO}(\vc{r}) \bar \psi_k(\vc{r}^\prime). 
\label{dp}
\end{equation}
In order to find the best GEOs, we minimize the Euclidean distance
$d=|U_\textrm{proj}-{\bar U_\textrm{proj}}|$ between the original and
the truncated geminal functions.
These functions are  defined in $ {\cal R}^3 \times {\cal R}^3$
in such a way that:
\begin{eqnarray} 
d^2  & = & |U_\textrm{proj}|^2 -2 \sum_k \int \!\! d\vc{r} d\vc{r}^\prime U_\textrm{proj}(\vc{r},\vc{r}^\prime) \phi^{GEO}_k (\vc{r}) 
\bar \psi_k( \vc{r}^\prime) \nonumber \\
& + &  \sum_k \int \!\! d\vc{r} ~ {\bar \psi}^2_k(\vc{r}), 
\label{expd2}
\end{eqnarray}
where $|U_\textrm{proj}|^2 = \int \!\!  d\vc{r} d\vc{r}^\prime
U^2_\textrm{proj}(\vc{r},\vc{r}^\prime)$, and we assumed that  the optimal atomic orbitals 
are orthonormal. This assumption is without loss of generality, because 
- whatever is the solution for the minimum -  we can always 
orthogonalize the corresponding 
 optimal orbitals $\phi_i^{GEO}$ and get a solution   
written in the same form as in Eq.(\ref{dp}). 
We can then take the variation over all possible unconstrained functions 
$\bar \psi (\vc{r})$ and show that the steady condition $ {\delta d^2
  \over \delta \bar \psi_k (\vc{r})} =0$ implies:
\begin{equation}
\bar \psi_k(\vc{r}) = \int \!\! d\vc{r}^\prime 
U_\textrm{proj}(\vc{r}^\prime,\vc{r}) \phi^{GEO}_k (\vc{r}^\prime).
\label{unconstrained_psi} 
\end{equation}
Replacing Eq.(\ref{unconstrained_psi}) into (\ref{expd2}) 
yields:
\begin{equation}
d^2= |U_\textrm{proj}|^2 - \sum_k  \int \!\! d\vc{r} d\vc{r}^\prime  D_\textrm{proj} ( \vc{r},\vc{r}^\prime) \phi_k^{GEO} (\vc{r}) 
\phi_k^{GEO} (\vc{r}^\prime), 
\label{new_d2}
\end{equation}
where $D_\textrm{proj}$ is the density matrix we have defined in Eq.(\ref{truedm}).
Thus, in order to minimize $d^2$ one needs to maximize the quadratic
form involving $D_\textrm{proj}$, with the constraint that the orbitals $\phi_k^{GEO} (\vc{r})$ 
are orthonormal. Therefore by the minimum/maximum  property of
symmetric operators (such as the positive definite
density matrix), it is clear that 
$d^2$ is minimized just when the optimal GEO orbitals coincide with the $p$ 
 eigenvectors of the density matrix with maximum eigenvalues $w_i$.
Indeed, all the  eigenvalues $w_i$  must be positive, and the
corresponding eigenvectors are obviously 
an orthonormal set of states, consistently with the assumption. 

From Eq.(\ref{truedm}) and the choice of the atomic projectors, it
follows that the density matrix kernel 
$D_\textrm{proj}$ can be expressed in terms of the 
atomic basis $\{\chi^\textrm{det}_i\}$ restricted around a given
atom at the selected position $\vc{R}_i=\vc{R}$.
By consequence, also the optimal GEOs can be expanded on the same local basis:
\begin{equation}
 \phi^{GEO}_i (\vc{r}) = \sum\limits_{j| \vc{R}_j=\vc{R}} \mu^{GEO}_{ij} \chi^\textrm{det}_j(\vc{r}). 
\label{geo}
\end{equation}
In the non-orthogonal finite basis $\{\chi^\textrm{det}_j\}$, this turns into  
the generalized eigenvalue Equation:
\begin{equation}
\sum\limits_{j| \vc{R}_j=\vc{R}} \left[( \lambda s \lambda^{\dag} ) s\right]_{ij} \mu^{GEO}_{kj} = w_k 
\mu^{GEO}_{ki} ~~~~\textrm{for $i$ s.t. $\vc{R}_i=\vc{R}$},
\label{solver}
\end{equation}
where the matrix $\lambda$ has been defined through Eq.(\ref{proj_det_mat_short}).
Eq.(\ref{solver}) can be immediately solved by standard linear 
algebra packages \footnote{This linear problem corresponds to a
  generalized eigensystem equation of the type $ A B x = \lambda x $
  with $A = \lambda s \lambda^\dag$ a symmetric matrix and $B=s$ a
  symmetric and positive definite one, $x$ and $\lambda$ being the
  eigenvector and the corresponding eigenvalue, respectively.}, by
considering that the overlap matrix $s$ is symmetric and  
positive definite. 
After diagonalization the eigenvector coefficients satisfy 
the orthogonality requirement  $\mu^{GEO} s (\mu^{GEO})^{\dag}=I$, 
that we have previously assumed. Moreover, the truncation error, i.e. the
residual distance, is $d^2 = |U_\textrm{proj}|^2 - \sum_{i=1}^p w_i$.

\subsubsection{GEOs properties}
\label{geos_properties}

Because the $\{ \phi^{GEO}_i \}$ basis set is optimal in the sense
defined in Secs.~\ref{general_framework} and \ref{detailed_procedure},
it has the advantage of not only 
being the best compromise between size and
accuracy, but also carrying the physical
information on the most representative atomic states for a site
embedded and interacting with its environment.  

Indeed, we notice that according to Eq.(\ref{geo})
the best GEOs are hybrid orbitals, as they are expanded over the full
set of atomic angular momenta. Thus, they can take care of
non trivial chemical hybridizations and, for instance, the crystal
field effect in solids is also automatically taken into account, as we
will see in Sec.~\ref{cerium}.

In the case of a starting geminal representing a Slater determinant, 
after the determination of the optimal basis $\{ \phi^{GEO}_i \}$ one can rewrite 
the same Slater determinant in Eq.(\ref{JSD}), within a target
accuracy, by expressing all the molecular orbitals in terms of the GEO
basis and with a dramatic reduction of the basis dimension 
and the number of variational parameters.

As introduced in Sec.~\ref{general_framework}, 
our embedding scheme naturally deals with a
determinantal part not necessarily restricted to a SD
form. Indeed, as the best GEOs are obtained by a distance minimization
in the $ {\cal R}^3 \times {\cal R}^3$ space, this can be applied not
only to geminals of Slater type,
but also to more generalized types with an arbitrary number of MOs $(M \ge N/2)$, and with
free parameters $\{ \lambda_i^\textrm{AGP} \}$. This would
correspond to the AGP form in Eq.(\ref{pairing_MO}), as explained in
Sec.~\ref{wf}. While for Slater determinants a relation can be found
with the NHOs (see Sec.~\ref{relation_to_nho}), the more
general AGP is less trivial, but it is rigorously accounted for by the
formalism presented in this work, which is generally applicable to \emph{any}
form of geminal functions. Therefore, via this scheme one can find a
GEO basis set which is optimal for both SD and AGP types of wave functions.

\subsubsection{Choice of the atomic projectors}
\label{atom_proj_choice}

In the geminal embedding method, 
we use Eq.(\ref{geo_proj}) to single out the local states at $\vc{R}$
in the left-projected geminal of Eq.(\ref{universe}).
The one in Eq.(\ref{geo_proj}) is not the only
possible way to define a projection around an atomic center $\vc{R}$. 
For instance, one could have
used the ``standard'' atomic projectors
\begin{equation}
\mathcal{P}^\textrm{at}_\vc{R}=\sum\limits_{i | \vc{R}_i=\vc{R}}
\sum\limits_{j | \vc{R}_j=\vc{R}} | \chi^\textrm{det}_i \rangle
s_{i,j}^{-1} \langle \chi^\textrm{det}_j |,
\label{atomic_proj}
\end{equation}
with $s$ the overlap matrix,
and defined $\psi^\textrm{proj}_{i,\vc{R}} =  \mathcal{P}^\textrm{at}_\vc{R}
\psi^{MO}_i$. However, we found that the optimal choice is the one of
taking only the basis set elements centered
on the atom $\vc{R}$ in Eq.(\ref{MO_expansion}), therefore defining $\psi^\textrm{proj}_{i,\vc{R}} = 
\sum\limits_{j | \vc{R}_j=\vc{R}} \mu_{ij}\chi^\textrm{det}_j$, as in Eq.(\ref{geo_proj}).

In order to understand this property, just consider the case  
when the basis is redundant around the given atomic center $\vc{R}$
(complete in $\vc{R}$, overcomplete when 
all atomic centers are taken into account). This  implies 
that Eq.(\ref{universe}) can be expanded in a much smaller number
of elements, and this refinement of the basis around $\vc{R}$ can be
easily determined by   
focusing on $U_\textrm{proj}^\vc{R}({\bf r},{\bf r^\prime})$, as it
has been described in Sec.~\ref{general_framework}.
Instead, if we use the 
projector $\mathcal{P}^\textrm{at}_\vc{R}$, this 
becomes the identity in the limit of large basis and it is therefore
not possible to disentangle a better localized basis set.  

In the example mentioned above, namely in the case of a redundant
basis and by using the projector in Eq.(\ref{geo_proj}), 
we can still describe Eq.(\ref{universe})
in terms of only $p<M_A$ appropriate atomic orbitals,  
given by  the eigenvectors of  $D_\textrm{proj}$ 
corresponding  to its $p$ non-zero eigenvalues. 
On the other hand, if we use this criterion of   
 basis reduction for all the atomic positions $\vc{R}_i$, we obtain the full 
geminal $\phi(\vc{r},\vc{r}^\prime) =
 \sum_\vc{R} U^\textrm{proj}_\vc{R}({\bf r},{\bf r^\prime})$  {\em 
exactly}. In the practical implementation (Sec.~\ref{detailed_procedure}), 
we remove also eigenvectors 
of $D_\textrm{proj}$ with small eigenvalues, by making therefore 
an approximation. It is clear however that this approximation 
can be systematically  controlled by decreasing the threshold of the 
accepted eigenvalues of $D_\textrm{proj}$.

\subsubsection{Relation with standard natural hybrid orbitals}
\label{relation_to_nho}

The ``standard'' atomic natural hybrid orbitals (NHOs) are defined as
eigenstates of the \emph{local atomic density matrix}
$D^\textrm{atomic}$, both left and right projected on a given
site.

The density matrix $D(\vc{r},\vc{r}^\prime)$ in general notations is a
two-point correlation function that, for wave functions in the SD representation,
coincides with the geminal $\phi(\vc{r},\vc{r}^\prime)$,
by setting $M=N/2$ and
$\lambda^\textrm{AGP}_i=1$ for $i=1,\ldots, M$ in
Eq.(\ref{pairing_MO}). Equivalently, the 
density matrix coincides with our unrestricted expression of $U$ in
Eq.(\ref{universe}), namely 
obtained with $\psi^\textrm{proj}_i=\psi^{MO}_i$. In the same SD limit,
$D=U=U^2$, for orthonormal MOs. 

Therefore, in order to determine the NHOs, one needs to define the
local atomic density matrix as
$D^\textrm{atomic}_{\vc{R}}=\mathcal{P}^\textrm{at}_\vc{R} D
\mathcal{P}^\textrm{at}_\vc{R}$. In the SD limit,
this is equivalent to
\begin{equation}
D^\textrm{atomic}_{ij}=\langle \chi^\textrm{det}_i | \phi |
\chi^\textrm{det}_j \rangle ~~~~\textrm{for $i$ s.t. $\vc{R}_i=\vc{R}$, and $j$ s.t. $\vc{R}_j=\vc{R}$}.
\label{standard_dm}
\end{equation}
This is clearly different from
our definition of projected density matrix $D_\textrm{proj}$ in
Eq.(\ref{truedm}), as in the latter case we do not use  
the standard atomic projection operators.

The second important difference takes place for wave functions 
beyond the SD representation, 
when the geminal $\phi$ is no longer
equivalent to the density matrix $D$. Therefore, also
$D_\textrm{proj}$, based on atomic
projected $\phi$, will differ from $D^\textrm{atomic}$, based
on atomic projected $D$, no matter what the atomic projector is.

We will compare the GEOs generated by the scheme in
Sec.~\ref{detailed_procedure} with the NHOs
obtained as solution of the following linear system:
\begin{equation}
\sum\limits_{j| \vc{R}_j=\vc{R}}  D^\textrm{atomic}_{ij}
\mu^\textrm{NHO}_{kj} = w_k \sum\limits_{j| \vc{R}_j=\vc{R}}  s_{ij}
\mu^\textrm{NHO}_{kj}
~~~~\textrm{for $i$ s.t. $\vc{R}_i=\vc{R}$},
\label{solver_nho}
\end{equation}
by taking the first $p$ NHOs with the largest $w_k$:
\begin{equation}
 \phi^{NHO}_k (\vc{r}) = \sum\limits_{j| \vc{R}_j=\vc{R}}
 \mu^\textrm{NHO}_{kj} \chi^\textrm{det}_j(\vc{r}).  
\label{nho}
\end{equation}
This choice
corresponds to the ``maximum-occupancy''
orbitals\cite{natural_hybrids}. 
In Sec.~\ref{water} and for SD wave functions, when a
direct comparison between GEOs and NHOs is possible, 
we will show that our embedding scheme yields GEOs that
provide much better performances in reducing the size of the atomic basis, 
compared to the standard NHOs discussed above.

\section{Selected applications}
\label{results}

In this section we illustrate the 
relevance of our embedding scheme 
by showing three examples: the
determination of the best GEO basis
set in water (Sec.~\ref{water}), cerium (Sec.~\ref{cerium}), and
hydrogen (Sec.~\ref{hydrogen}).
In the first case, the natural orbitals determined by the density
matrix embedding method significantly reduce the number of wave function
parameters and so its computational burden. In the second case, the
GEO basis set carries the physical 
information on the atomic structure of a cerium site embedded in the crystal
environment, and allows a physical interpretation of the electronic
change underlying the $\alpha$-to-$\gamma$ volume collapse. In the
last example, the GEO basis is shown to be the best compromise
between accuracy and efficiency in the determination of the phase diagram
of liquid hydrogen at high pressure by means of QMC-based molecular
dynamics (MD) calculations, which require both accurate and cheap wave
functions. 

\subsection{Water molecule}
\label{water}

A proper description of the water molecule is essential to reproduce
the structural and electronic properties of larger water cluster,
liquid water and ice, because in large systems containing several
molecules there is a strong interplay between the intramolecular and
intermolecular degrees of freedom, due to the large water dipole
moment and the strong directionality of the H bond. For this reason,
the water molecule has been the subject of many theoretical
works\cite{fink,exact_H2O_energy,water_needs,sorella_hb,alfe},
aiming at finding the quantum chemistry method which has the best
balance between accuracy and computational cost.

In this regard, QMC methods are promising, thanks to their favorable
scaling with the system size. Here we report our pseudopotential and
all-electron calculations\cite{mario_water} on 
the water molecule with different wave
function types and basis sets, in order to show that the speed-up
offered by the GEOs basis set is relevant, and can open the way to
more systematic studies on larger water systems.

\begin{table*}[!ht]
  \resizebox{0.95\textwidth}{!}{\begin{minipage}{\textwidth}
\caption {VMC energies of the water molecule and number of variational
  parameters in the QMC wave functions.
  The geometry is the experimental one in the
  pseudopotential calculations, while is the QMC relaxed one in
  all-electron calculations. See Ref.~\onlinecite{mario_water} for more
  details. The total number of parameters (last column) and the wave function
  quality vary depending on the contraction level of the GEOs used in
  the determinantal part.   
  The Jastrow functional form has been kept fixed in all set of calculations.
  This gives a number of 195 and 418 Jastrow parameters for
  the pseudopotential and all-electron calculations, respectively.
  The other parameters are in the determinant, coming from both
  $\lambda^{a,b}$ (third to last column) and the basis set,
  i.e., $\chi^\textrm{det}_a$ for the primitive GTO and $\phi^{GEO}_a$ for
  the GEOs (second to last column).} \label{table_water_energy} 
\begin{tabular}{|r|d|d|d|r|r|r|}
\hline
  \multicolumn{1}{|r|}{Wave function ansatz}   & 
\multicolumn{3}{c|}{VMC energies} &
\multicolumn{3}{c|}{number of parameters} \\ 
\multicolumn{1}{|c|}{} & 
   \multicolumn{1}{c|}{Energy $E_x$ (Ha)} &
   \multicolumn{1}{c|}{Variance (Ha$^2$)}   &  \multicolumn{1}{c|}{$E_x
     - E_\textrm{JSD}$ (mHa)} &
   \multicolumn{1}{c|}{$\lambda^{a,b}$} &
   \multicolumn{1}{c|}{$\chi^\textrm{det}_a, \phi^{GEO}_a$} & \multicolumn{1}{c|}{total} \\
\hline
\multicolumn{7}{|l|}{pseudopotential calculations} \\
\hline
JSD:  primitive GTOs   &             -17.24821(7)      &
0.2655(6) &  0.0 & 682 & 18 & 895\textsuperscript{\emph{a}}\\ 
JAGP: (4O,1H) GEOs  &   -17.25013(8)      & 0.2635(12)  &
-1.91(11)  & 21 & 158 & 374\\ 
JAGP: (4O,5H) GEOs  &   -17.25183(6)      &  0.2510(6) &
-3.62(10) & 105 & 238 & 538 \\
JAGP: (8O,2H) GEOs  &   -17.25267(7)     &   0.2426(18) &
-4.46(10) & 78 & 298 & 571 \\
JAGP: (8O,5H) GEOs  &   -17.25302(6)     &   0.2412(34) &
-4.89(10) & 171 & 358 & 724 \\
JAGP:  primitive GTOs    &  -17.25389(6)   &  0.2296(5) &
-5.68(10) & 682 & 18 & 895 \\
\hline
\multicolumn{7}{|l|}{all-electron calculations} \\
\hline
JSD:    primitive GTOs   &             -76.40025(8)      &
1.412(3) &  0.0 & 1383 & 19 & 1820\textsuperscript{\emph{a}}\\ 
JAGP: (9O,2H) GEOs  &   -76.40504(9)    &   1.399(6)  &
-4.79(12) & 91 & 361 & 870 \\
JAGP: primitive GTOs    &  -76.40660(7)   &  1.374(3) &
-6.35(11) & 1383 & 19 & 1820 \\
\hline
\end{tabular}
\footnotetext[1]{Here the number of parameters is the same as the one
  in the JAGP wave function since in the JSD ansatz
  we rewrite the corresponding geminal (of rank $N/2$) on
  the uncontracted basis in order to optimize the MO's, as explained by
  \citet{michele_agp2}.}
\end{minipage}}
\end{table*}

The BFD pseudopotential\cite{filippi_pseudo} has
been used for oxygen, while the two hydrogens have been treated
all-electron. We have also performed full all-electron calculations,
for both oxygen and hydrogen. The primitive Gaussian basis set for
oxygen is $(5s,5p,2d)$ and $(6s,6p,2d)$ in pseudopotential and
all-electron calculations, respectively. For hydrogen, the primitive
basis set is $(4s,2p)$. The Jastrow functional form has been kept fixed
and developed on a primitive Gaussian basis set of $(3s,2p,1d)$ and
$(2s,1p)$ for oxygen and hydrogen, respectively. 
Note that this GTO set has been recently claimed to
be one of the most accurate in an extensive 
QMC study of single molecule water properties\cite{zen_water}, which
used the same Jastrow ansatz as ours. 
For the antisymmetric part we tested two main wave
function forms, the single Slater determinant (obtained by using a
geminal with rank $N/2$), and the AGP function. 
At variance with Eq.(\ref{pairing_MO}), for the water molecule we chose
to develop the AGP geminal directly on the primitive GTOs (and not on
MOs) to have a greater flexibility, such that: 
\begin{equation}
\phi({\bf r},{\bf r}^\prime)= \sum_{a,b=1}^{M_A \times N_\textrm{atoms}} \lambda^{a,b}
\chi^\textrm{det}_a (\textbf{r}) \chi^\textrm{det}_b (\textbf{r}^\prime).
\label{GTO_expansion}
\end{equation}
The energy difference between the JSD and the JAGP wave functions, reported in
Tab.~\ref{table_water_energy}, shows the size of static
correlations in the system, which amounts to 5-6 mH. Moreover, the AGP
ansatz provides a better description of the nodal surface, because
lattice regularized diffusion Monte Carlo (LRDMC)
calculations\cite{lrdmc,filippi_lrdmc} give a fixed-node energy which is 2.5 mHa
lower than the one obtained by using the JSD trial wave function\cite{mario_water}.
The JAGP wave function leads also to better geometrical
properties\cite{mario_water}. Its relaxed geometry is closer to the
experiment than the JSD one, in both the OH distance and the HOH
angle. 

\begin{figure}[h]
\includegraphics[width=\columnwidth]{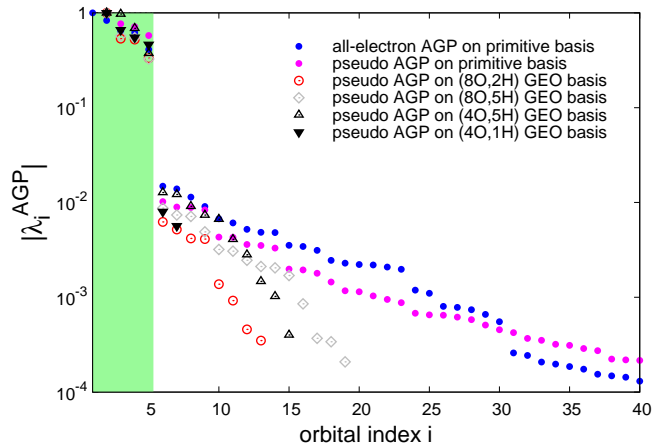}
\caption{Semilog plot of the modulus of the AGP eigenvalues
  versus the MO index for different basis sets
  and calculations. The orbital indexes include always the oxygen 1s
  electrons, replaced in the pseudopotential calculations.
  The green area represents the exactly occupied molecular
  orbitals in the single Slater determinant representation, with $\lambda_i=1$
  for $i \in \{ 1, \ldots, \textrm{HOMO} \}$ and $\lambda_i=0$ for $i
  \ge \textrm{LUMO}$. 
  In the AGP, also the orbitals above the HOMO are occupied, with a
  weight $|\lambda^\textrm{AGP}_i|$ which
  jumps at the HOMO-LUMO level. The AGP developed on the GEO
  basis set correctly reproduces the HOMO-LUMO jump and spans the same
  relevant region in the AGP eigenvalues spectrum.
Figure adapted from Ref.~\onlinecite{mario_water}.
} \label{AGP_eigenvalues}
\end{figure}

To analyze how the AGP correlations develop in the water molecule, we
diagonalize the geminal of Eq.(\ref{GTO_expansion}) in order to recast
it in its MOs representation of Eq.(\ref{pairing_MO}). Indeed,
the diagonalization of $\phi$ in the space spanned by the
basis set $\chi^\textrm{det}_a$ yields
the MOs as eigenvectors and $\lambda^\textrm{AGP}_i$ as
eigenvalues, whose absolute values are 
plotted in Fig.~\ref{AGP_eigenvalues}. 

Fig.~\ref{AGP_eigenvalues} shows that indeed the orbitals
above the HOMO have a sizable weight, with a
distribution which falls abruptly to zero only after the 40-th
orbital (not reported in the Figure). This reflects the multi determinant
character of the water molecule, taken into account by the AGP
ansatz. 

We turn now the attention on how to reduce the AGP basis set in an
effective way. So far, both the JSD and JAGP wave functions have been
developed on the primitive basis in order to exploit at most its
flexibility. Thus, the total number of variational parameters is
895 in pseudopotential calculations (see the last column of
Tab.~\ref{table_water_energy}), 
quite large for a single molecule, particularly if one would like to tackle 
the study of larger water clusters by means of QMC techniques. 
The most important limitation of this approach is that the number of
variational parameters corresponding to the matrix elements $\lambda^{a,b}$
increases as the {\rm square} of the atomic basis size. Therefore this
should be reduced at minimum in order to make this approach feasible 
for a large number of molecules.

To this purpose, we would like to find the optimal contracted basis
set. We start from our best JAGP wave function previously optimized,
we take its Jastrow factor off, and we are left with its geminal part
of Eq.(\ref{GTO_expansion}). From there, we follow the recipe explained
in Sec.~\ref{method}, and we generate a new set of GEOs, upon which we
develop a new geminal:
\begin{equation}
\tilde \phi({\bf r},{\bf r}^\prime)= \sum_{a,b=1}^{p 
\times N_\textrm{atom}} \tilde \lambda^{a,b}
\phi^{GEO}_a (\textbf{r}) \phi^{GEO}_b (\textbf{r}^\prime),
\label{GEO_expansion}
\end{equation}
where the $\tilde \lambda^{a,b}$ are given by overlap maximization of
the latter $\tilde \phi$ with the original $\phi$ in the $ {\cal R}^3
\times {\cal R}^3$ space. Thanks to the GEO expansion, the overlap is 
supposed to be good even for small GEO basis set size $p$. 
We found that, at fixed $p$, the normalized overlap $\langle \tilde\phi | \phi
\rangle^2/(\langle\phi | \phi \rangle \langle \tilde\phi | \tilde\phi
\rangle)$ is systematically larger when one uses the GEOs obtained as in
Sec.~\ref{embedding_scheme} than the NHOs obtained by
diagonalization of the atomic (left \emph{and} right projected) $U^\textrm{atomic}$
matrix (the ``maximum-occupancy'' NHOs\cite{natural_hybrids} of Eq.~\ref{nho}).
The overlap between the geminal function $\phi$ developed in the primitive
basis and the one ($\tilde \phi$) written in the GEOs basis sets is reported in
Tab.~\ref{overlap}. For the case in latter Table, the starting $\phi$ is the optimal one for the
JSD energy of the water molecule in the primitive basis. 
Our embedding scheme systematically gives GEOs which yield a better
overlap between $\phi$ and $\tilde \phi$, at a fixed GEO size, if
compared with the overlaps obtained with the ``standard'' NHO embedding
scheme. 

\begin{table}[htp]
\begin{tabular}{| l | d | d |} 
\hline 
 \multicolumn{1}{|c|}{GEO/NHO} & \multicolumn{1}{|c|}{GEOs} & \multicolumn{1}{|c|}{``standard''
  NHOs} \\ 
 \multicolumn{1}{|c|}{basis set} & \multicolumn{1}{|c|}{overlap
   ($\%$)} & \multicolumn{1}{|c|}{overlap ($\%$)} \\ 
\hline 
(4O,1H)  &           99.8390  &      99.0696 \\
(8O, 2H) &           99.9511 &       99.1846 \\
(4O, 4H) &         100.0000 &       99.1929 \\
(4O, 5H)  &        100.0000   &     99.1933  \\
(8O, 5H) &         100.0000  &      99.2188 \\
(12O, 6H) &       100.0000 &       99.8305 \\
(20O, 8H) &       100.0000   &     99.9458 \\
(30O, 10H) &     100.0000 &      100.0000  \\
\hline 
\end{tabular} 
\caption{Overlap $\langle \tilde\phi | \phi
  \rangle^2/(\langle\phi | \phi \rangle \langle \tilde\phi | \tilde\phi
  \rangle)$ between the geminal $\phi$ of the fully optimized
JSD wave function in the primitive basis set and the best  $\tilde
  \phi$ developed on the GEO/NHO basis set reported
  in the first column (with $\tilde \lambda^{a,b}$ which maximize the $ {\cal R}^3
  \times {\cal R}^3$ overlap). Given the basis set size, the
  contracted atomic orbitals are determined
  in the ``standard'' way (third column, see Eqs.~\ref{solver_nho} and
  \ref{nho} for definition) and
  by the geminal embedding scheme described in Sec.~\ref{detailed_procedure}
  (second column). The embedding 
  scheme presented here systematically gives better overlaps. 
 The GEOs converge to full overlap already for a (4O, 4H) basis set,
 as the dimension of the ``universe'' is 4 in an SD wave function (the total number of electrons
 is 8 plus the $1s^2$ core electrons of oxygen), and in the GEO
 embedding framework the maximum number of local states per nucleus is at most
 the dimension of the ``universe''.
The last line corresponds to the complete basis set limit for the
contractions with respect to the space spanned by the primitive basis
set, where all methods have to converge by definition (as it is
actually found numerically). 
  \label{overlap}}
\end{table}

From Eq.(\ref{GEO_expansion}), it is apparent that the reduction of variational parameters
with respect to the expression in Eq.(\ref{GTO_expansion}) is a factor
$p^2/M_A^2$, which can be extremely relevant if $p \ll M_A$. In the
following, we are going to study the VMC energy
convergence of
the JAGP wave function expanded in GEOs as a function of $p$,
by comparing it with its rigorous lowest energy limit provided by the
uncontracted JAGP reference previously computed. We recall that once
the GEOs expanded AGP is obtained, its parameters are further
optimized in the presence of the Jastrow factor by QMC energy
minimization, in order to find the best variational wave function
within the JAGP ansatz in the GEO basis set.

The energy results are reported in Tab.~\ref{table_water_energy}. 
For the pseudopotential calculations, 
the (4O,1H) GEO basis set ($p=4$ for oxygen, $p=1$ for hydrogen)
is the smallest basis set which can take into account the $2s2p$ near
degeneracy at the atomic O level, including the 1s for H and the
$2s$ and $2p$ orbitals for O. Its energy and variance are the poorest 
among the GEO basis sets considered in the Table, though being lower than the JSD
ansatz. 
On the other hand, 
the largest GEO basis used here, namely the (8O,5H)
set, recovers a large fraction of static correlation and its energy is
less than 1mHa above the uncontracted JAGP one. 
However, the parameter reduction is weak (see
last column of Tab.~\ref{table_water_energy}), with a total of 18 GEO basis set
elements against 50 elements (counting also the azimuthal multeplicity) in
the corresponding primitive basis set. 
The best compromise between efficiency, i.e. total number of
variational parameters, and accuracy, i.e. variational energy, is provided by the
(8O,2H) basis, as it yields a significant gain in energy
with a small/moderate number of parameters.

The same behavior has been confirmed in all-electron calculations,
reported in the lower panel of Tab.~\ref{table_water_energy}, where
the 1s electrons are explicitly kept in the oxygen atom. The
corresponding GEO basis set has been extended to 9O GEO orbitals,
performing in the same way as in the case of pseudopotential
calculations in the 8O GEO basis set. Even in this case, the gain in
the number of variational parameters is important, without a
significant deterioration of the wave function with respect to the
JAGP in primitive basis.

Finally, we study how the AGP spectrum changes with the 
contracted GEO basis sets. Fig.~\ref{AGP_eigenvalues} shows that, after a
complete wave function optimization, the natural orbital eigenvalues
magnitude of the
GEO AGP covers the $10^{-2}-10^{-4}$ range of
the primitive AGP, except for the shortest (4O,1H)
basis, which clearly spans a too small Hilbert space. Moreover,
we checked that the JAGP expanded on the optimal (8O,2H)
basis gives the same fixed node LRDMC energy as the full JAGP,
signaling that the nodal surface is properly described even by the
(8O,2H) GEO contraction.

The advantage of using the GEO basis set will be remarkable for
larger systems with many water molecules,  
as the number of variational parameters corresponding to the 
GEO orbitals grows only \emph{linearly} with the 
number of atoms. Instead the number of parameters corresponding 
to $\tilde\lambda^{a,b}$,
grows quadratically, but it remains still affordable since 
it is dramatically reduced by this approach.

\subsection{Solid cerium}
\label{cerium}

\begin{figure}[h]
\includegraphics[width=\columnwidth]{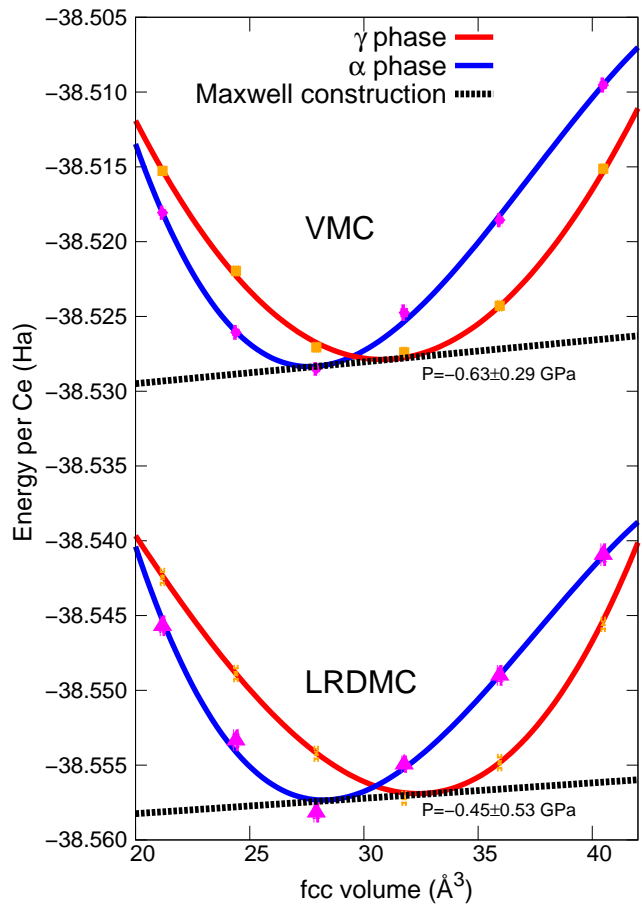}
\caption{Equation of state of the $\alpha$ (blue line) and $\gamma$
  (red line) phases of
  elemental cerium computed by VMC and LRDMC. In the
  y-axis we report energies per atom of the pseudo
  Hamiltonian. Also the common tangent construction is plotted (dashed
  black line),
  where the transition pressure is derived from. The experimental
  transition pressure extrapolated to 0 $T$ is $\approx -1$ GPa.  
\label{cerium_eos}}
\end{figure}

The $\alpha$-to-$\gamma$ phase transition in cerium, also known as
volume collapse, is one of the most challenging phase transitions in
Nature, as it is driven by the f-electron correlation, very hard to
treat by any many-body method. Cerium peculiar features, such as its
isostructural volume jump between two \emph{fcc} structures, are not
completely explained yet, and there is a strong debate whether
the $\alpha$-to-$\gamma$ transition would still be present at zero
temperature\cite{Svane1994,PhysRevLett.109.146402,PhysRevLett.111.196801,PhysRevB.91.125148,PhysRevB.91.161103}
and negative  
pressures (in the actual material the transition occurs only at
finite T\cite{decremps}), or there is a lower critical point of the first-order
transition line with $T_\textrm{min} > 0$\cite{PhysRevLett.49.1106}. The question is not
just academic, as the presence of a $T=0$ first-order transition implies
a purely electronic mechanism underlying the volume collapse, while
temperature effects are certainly due also to electronic entropy contributions which
are known to be important in f-electron
systems\cite{PhysRevB.67.075108,PhysRevLett.96.066402}. 
Very recently\cite{casulacerium}, accurate QMC
calculations show that a first-order transition still exists even at
$0 K$. This is reported in Fig.~\ref{cerium_eos}, where the equation of
state $E=E(V)$ is plotted for both variational Monte Carlo (VMC) and lattice
regularized diffusion Monte Carlo (LRDMC) calculations. 

As explained
in Ref.~\onlinecite{casulacerium}, the two solutions found at fixed
unit cell volumes, in a range compatible with the $\alpha$ and 
$\gamma$ phases, mainly differ for their Slater determinants, which are
fully optimized at the VMC level in the presence of the Jastrow factor
by energy minimization. 
Their MOs are developed on a large $7s7p4d5f1g$ GTO primitive basis set, as
in Eq.~\ref{MO_expansion}. From 
Fig.~\ref{cerium_eos}, it is apparent that the LRDMC further lowers
the variational energy of the state, although without changing the
physical picture of the transition. Indeed, the transition pressure and the
critical volumes are in agreement with the experimental ones
already at the VMC level. Interestingly enough, the nodal structure of
the two phases, set by the Slater determinant, is 
certainly different, as the fixed node approximation is the only
constraint in the LRDMC calculations which prevents one phase to be
projected to the other. 

\begin{figure}[t]
\includegraphics[width=\columnwidth]{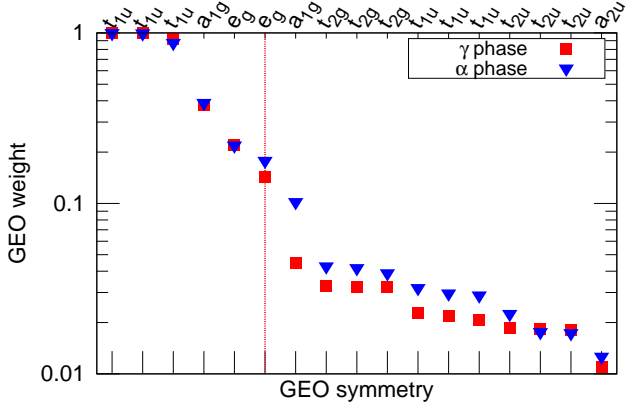}
\caption{Weights $w_i$ of the first 17 most relevant GEOs, computed by following the
  embedding scheme detailed in Sec.~\ref{embedding_scheme}. They are derived from
  the Slater determinants of the two wave functions describing the
  $\alpha$ (blue triangles) and $\gamma$ (red squares) phases, respectively,  at the unit cell volume of 31.7 \AA$^3$
  (fcc lattice space of 9.5 $a_0$). The vertical red line indicates the
  ``closed-shell'' occupation of the 12 electrons in the
  pseudopotential in the case of largely separated weakly interacting
  atoms.
Figure adapted from Ref.~\onlinecite{casulacerium}.
\label{ano_weights}}
\end{figure}

\begin{figure}[t]
\includegraphics[width=\columnwidth]{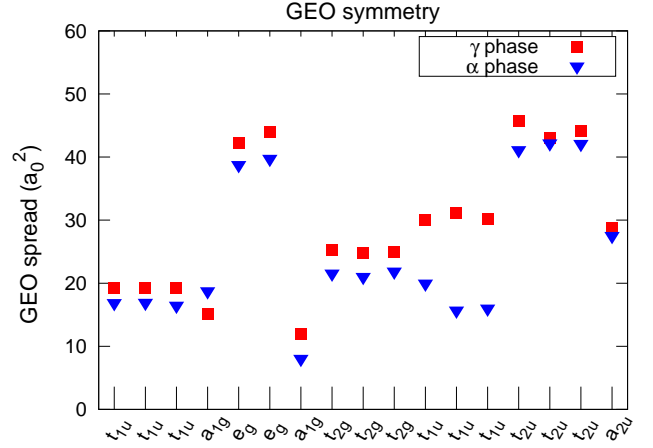}
\caption{Spread ($\langle \phi | r^2 | \phi \rangle - | \langle \phi | \vc{r}
  | \phi \rangle |^2$) of the same GEO basis as in
  Fig.~\ref{ano_weights}, for both the $\alpha$ (blue triangles) and
  $\gamma$ (red squares) phase. 
Figure adapted from Ref.~\onlinecite{casulacerium}.
\label{ano_spread}}
\end{figure}

In order to understand what is the electronic mechanism of
the volume collapse, an analysis of the Slater part of the JSD wave
function is thus necessary, as it bears information related to the electronic
structure change between the two phases. With this aim, the
embedding scheme proposed in the present paper is extremely useful, as the
derived GEOs, their weights (Fig.~\ref{ano_weights}) and their spread
(Fig.~\ref{ano_spread}) reveal the effect of the crystal field and the
hybridization on the underlying electronic structure. The GEO in this
case are naturally hybrid, as the crystal field mixes the atomic
components. Thanks to the
atomic orbital resolution of each GEO, it has been possible to assign
to each one its correct spatial symmetry, compatible with the fcc crystal point
group, which is reported in Figs.~\ref{ano_weights} and \ref{ano_spread}.

A feature which is clear from Fig.~\ref{ano_weights} is the tight
competition between many different atomic 
states, the s-based $a_\textrm{1g}$, the d-based $e_g$ and $t_{2g}$,
and the f-based $t_\textrm{1u}$,  $t_\textrm{2u}$, and
$a_\textrm{2u}$ orbitals. From band structure calculations, it is known that
they all have weight at the Fermi level, and they are very close in
energy. From our embedding
analysis, they all contribute to the total wave function with quite similar
weights $w_i$. This makes cerium a puzzling system, as its
physics is dictated not only by the strong correlation affecting the
f-orbitals but also by their complex interplay with more
delocalized $s$, $p$, and $d$ states.
Only the first 4 GEO orbitals (3 $t_\textrm{1u}$ and 1
$a_\textrm{1g}$) have weight close to 1, as they are almost perfectly
occupied by the $5s^2 5p^6$ semi-core electrons included in the HF energy consistent
pseudopotential\cite{dolg_private}, which are chemically inert. The
other 4 electrons in the pseudopotential go into the valence, and
occupy the higher energy GEOs according to an atomic participation
rate which is related to the GEOs $w_i$.
Fig.~\ref{ano_weights} shows that a main change between the
$\alpha$ and $\gamma$ phase is related to the weight of the
$a_\textrm{1g}$ (mainly $6s$) orbital, which reflects a different atomic
occupation and thus a different nature of
the chemical bond between neighboring atoms.

Another difference is apparent from Fig.~\ref{ano_spread}, where we plot
the GEOs spread. The most significant spread variation affects
the $t_\textrm{1u}$ orbitals, which are built upon a linear
combination of $p$ and $f$ atomic 
symmetries, allowed by the crystal field. The  $t_\textrm{1u}$,  $t_\textrm{2u}$, and
$a_\textrm{2u}$ GEO orbitals are the most correlated, as they are made
of $f$ atomic orbitals. A change in the $t_\textrm{1u}$ spread is a
strong indication that the correlation level in the two phases is very
different.

\begin{figure*}[t]
\includegraphics[width=2\columnwidth]{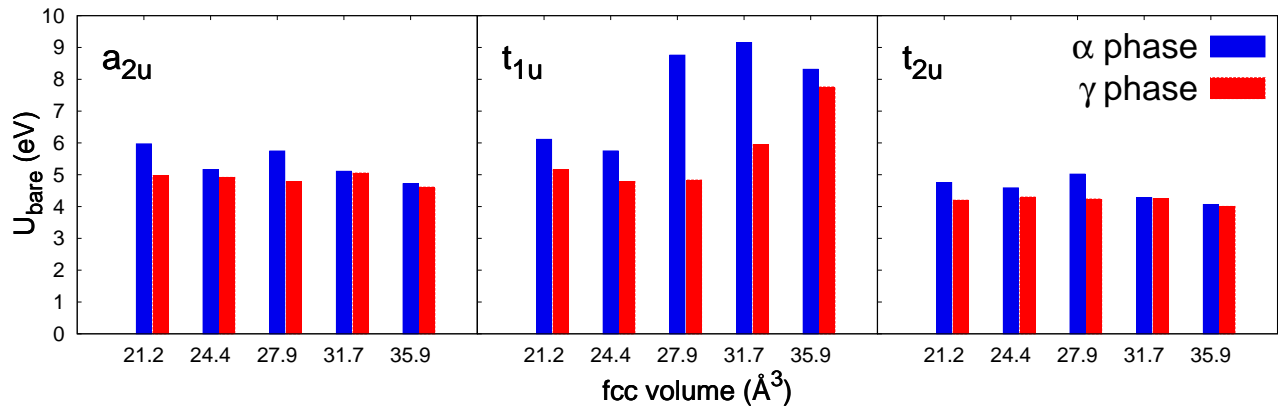}
\caption{Local Coulomb repulsion $U^k_\textrm{bare}=\langle \phi^\textrm{GEO}_k
  \phi^\textrm{GEO}_k | \frac{1}{|\mathbf{r}-\mathbf{r}^\prime|} |
\phi^\textrm{GEO}_k \phi^\textrm{GEO}_k \rangle$ computed on the GEO
basis set of $a_{2u}$, $t_{1u}$, and $t_{2u}$ symmetries, containing
the f-orbital components, for the $\alpha$ and $\gamma$ 
phases at different volumes (corresponding to the points in Fig.~\ref{cerium_eos}). 
\label{Uf}}
\end{figure*}

This is substantiated by the calculation of the local Hubbard
repulsion 
\\
$U^k_\textrm{bare}=\langle \phi^\textrm{GEO}_k
  \phi^\textrm{GEO}_k | \frac{1}{|\mathbf{r}-\mathbf{r}^\prime|} |
\phi^\textrm{GEO}_k \phi^\textrm{GEO}_k \rangle$, computed by using
the GEO orbitals as local atomic states, and plotted in
Fig.~\ref{Uf}. In accordance to the spreads in Fig.~\ref{ano_spread},
the main change in $U$ affects the $t_\textrm{1u}$ atomic states, with
the $\alpha$ phase which has a larger $U$ (at the \emph{fixed}
volume) than the $\gamma$ phase. This is rather counter-intuitive, as
the $\alpha$ phase is known to be less ``correlated'' than the
$\gamma$ one, as revealed for instance by a taller quasiparticle peak in
photoemission spectroscopy at the Fermi level. However,
these experimental signatures are obtained at different volumes, with
the $\alpha$ phase in its collapsed configuration ($15 \%$ volume less
than the $\gamma$) and so with an $U$-over-bandwidth ratio which is
larger in the $\gamma$ phase. In our analysis, the two phases can be
studied at the same volume, where two steady solutions are found.
This possibility allows one to unveil an important effect due to
the strong f-correlation: the $\gamma$ phase gains energy with respect
to the $\alpha$ phase, by reducing the local Coulomb repulsion in the
$t_\textrm{1u}$ channel, in the large-volume region where the
correlation is stronger for both phases. This is thanks to the breathing of the
$t_\textrm{1u}$  orbitals, strongly hybridized with the $p$ states,
and broader in the $\gamma$ phase. At its turn, this implies a larger
$t_\textrm{1u}-t_\textrm{1u}$  overlap between neighboring sites, and
so a stronger bonding character of the $t_\textrm{1u}$ orbitals in the
$\gamma$ phase, whereas in the $\alpha$ phase it is the
$a_\textrm{1g}$ ($s$) channel which has a stronger weight, as we have
already seen. 
According to molecular orbital theory, an $s$-based bonding molecular
orbital has a larger overlap and lower energy than the ones based on
degenerate (or almost degenerate) higher atomic angular
momenta. Therefore, the stronger $a_\textrm{1g}$ weight in the
$\alpha$ phase is compatible with a tighter chemical bond, and so a smaller
bond length and unit cell volume.  

To conclude, with our embedding analysis, we showed how 
on-site intraorbital repulsions in the $f$-$t_{1u}$ manyfold accompanied by
hybridization effects between the $f$ atomic orbitals with
more delocalized $s$ and $p$ states are responsible for the
stability of the $\gamma$ phase at large volumes, while the
$\alpha$ phase is more stable at smaller volume with shorter ($10-15 \%$
difference) lattice parameters.
The volume collapse
transition results therefore from a tight competition between  
interatomic bond strength and local Coulomb repulsion.

\subsection{Liquid hydrogen at high pressure }
\label{hydrogen}

In this Section we present recent results on
liquid hydrogen at high pressure\cite{secondmazz}, showing the  
advantage of the basis set reduction in this case.
We consider a molecular dynamics (MD) simulation within the BO
approximation, where at each step several thousands of parameters of a
JSD variational wave function have to be optimized. The use of the GEO
basis set, 1 GEO in the $2s$ GTO case, and 3 GEOs in the
$3s1p$ case, allows for a systematic  
reduction of  the number of variational parameters by a factor four as
compared with the primitive basis, without a sizable loss of
accuracy (within 1mHa/atom), as seen also in the case of the water
molecule (Sec.~\ref{water}).
The reduction of the basis is achieved 
at the beginning of the simulation, in order to start with the best
possible variational  
wave function compatible with the small basis set chosen.
The JSD wave function is composed of a Slater determinant, whose
molecular orbitals are expanded in the  
chosen basis and that can be obtained by a DFT calculation or by a
full optimization of the energy  
in presence of a Jastrow factor, as described in Sec.~\ref{wf}.
A DFT calculation is first performed in the primitive basis for typical 
configurations at different densities and fixed hydrogen positions. 
After that,  the contraction, namely the determination of the initial
optimal GEOs, is performed   with the algorithm described in
 Sec.~\ref{embedding_scheme}. Finally, the full optimization  
of the JSD is performed and the best JSD compatible with the GEO basis
set is obtained by means of  state-of-the-art optimization
techniques\cite{srh,sorella2007}.  
Thanks to the contraction, the number of parameters to be optimized
becomes affordable. 
Moreover the optimization is particularly fast and reliable 
due to the very good initial guess obtained in this way. 
That makes the search for the global minimum (or at least a very good
one) much simpler, being the complexity of the energy landscape reduced.
For these reasons in the largest primitive basis $(6s5p1d)$\cite{bfd}
the statistical optimization of the molecular orbitals  
is not implemented, because not possible or at least very difficult, and
only the Jastrow factor is optimized. In this case however, based on
smaller system calculations, 
we do not expect meaningful improvement in the energy by full
optimization, as the determinant  
obtained in a large basis is essentially optimal within DFT and the
LDA approximation.   
The results are summarized in Tab.~\ref{tableen} and compared with the
ones obtained with other techniques,  
among which LRDMC and Reptation Quantum Monte Carlo (RQMC), namely
different methods to obtain the lowest variational energy compatible
with the nodal surface of a given variational ansatz.  
LRDMC and RQMC values are obtained with standard extrapolations with
the mesh size or time step, respectively, as only in this limit they
coincide if the same wave function ansatz is used in both methods. 

\begin{table}[htp]
\begin{tabular}{l c c } 
\hline\hline 
 & \multicolumn{1}{c}{$r_s=1.44$} & \multicolumn{1}{c}{$r_s= 1.24$}
 \\ [0.5ex] 
\hline 
VMC - JSD - \emph{2s}/1GEO  & -0.54750(1) &  -0.5141(1) \\
VMC - JSD - \emph{3s1p}/3GEOs & -0.55391(1) &  -0.52246(1)\\
VMC - JSD - \emph{6s5p1d} primitive & -0.5542(1) &  -0.5230(1) \\
\hline 
 LRDMC - JSD - \emph{2s}/1GEO & -0.55239(1) &  -\\
 LRDMC - JSD - \emph{3s1p}/3GEOs & -0.55678(1) &  -0.52535(1)  \\ 
 \hline
 VMC -~ JSD+B & -0.55605(2) &  -0.52564(2)\\
 RQMC -~ JSD+B  &-0.5572(1) &   -0.52638(3)  \\ [1ex] 
\hline 
\end{tabular} 
\caption{ Total energies per atom (in Hartree) obtained for fixed
  ionic configurations at different $r_s$ with different methods: VMC,
  LRDMC or RQMC; with different basis sets as described in the
  text. JSD+B refers to the Jastrow-Slater wave function with backflow
  transformation.  
\label{tableen}}
\end{table}

As it is clear from Tab.~\ref{tableen}, by using GEOs a small basis is enough
to obtain almost converged results  
well below  1mHa/atom, if we take as a reference the result of the
largest primitive basis mentioned above.  
The best energies for the two densities (defined by $r_s$, such that
$\frac{4}{3}\pi r_s^3 a_0^3$ is the volume per particle, $a_0$ being the Bohr radius) reported here are obtained with the
JSD+backflow (JSD+B) ansatz, which is used in the coupled electron-ion
Monte Carlo (CEIMC) calculations\cite{Morales:2010p28600}. Here both
the determinant (SD) and  
the Jastrow (J) are obtained with a different approach 
 and with a different basis set. Moreover the determinantal part SD
 contains also  backflow correlations that are supposed to improve
 substantially the nodal surface of a simple  Slater determinant. 
Nevertheless  it is quite clear that the present  
variational approach also in a small basis and restricted to a 
 single determinant can provide reasonably accurate energies. 
Moreover the LRDMC applied to such guiding functions is very close 
(within  $\lesssim$ 1 mHa/atom) to the best RQMC
results\cite{PhysRevLett.82.4745},   
implying that the \emph{3s1p}/3GEOs basis provides an almost optimal
nodal surface. 

After these successful optimizations of the basis set we can perform
MD simulations because the number of variational parameters is
feasible, as at each iteration of MD we can  
optimize the wave function with few steps (about $6$) and compute at the
final iteration the atomic forces with a reasonable computational effort.
The interested reader can  find more details of the MD used 
 in our previous papers\cite{yeluo_draft_2014,firstmazz,secondmazz}, as
 this is outside the scope  of the present work.  

\begin{figure}[h]
\includegraphics[width=\columnwidth]{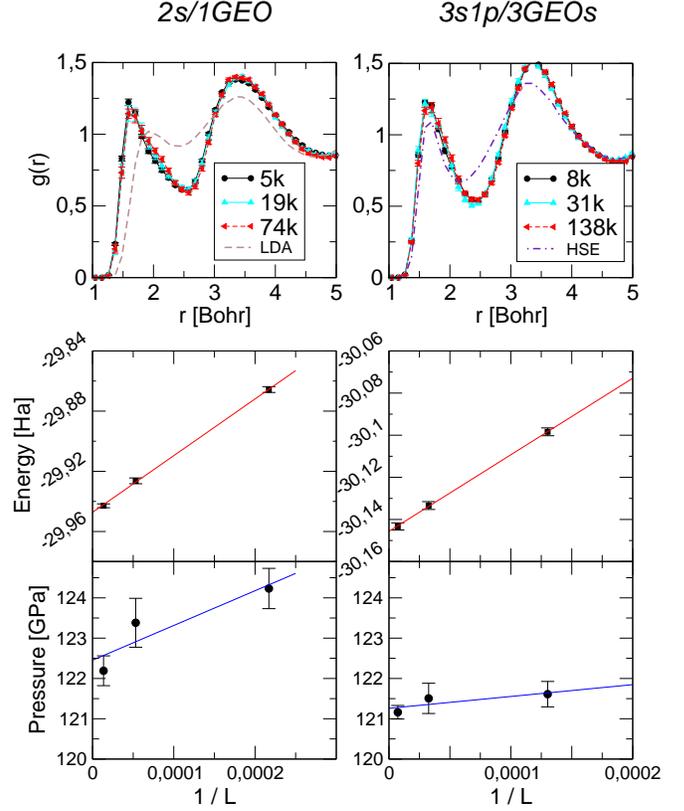}
\caption{Accuracy of the wave function: tests at density $r_s$=1.44 and T=600
K. The left panels refer to the $2s$/1GEO basis set, while the
right panels to the much more accurate $3s1p$/3GEOs.
\emph{Top panels:} Radial distribution function $g(r)$ obtained for
different $L$ (VMC sample size per MD iteration). 
We also plot the DFT predictions for LDA and
HSE xc functionals. All the calculations are performed for 54 atoms at
the  $\Gamma$ point. 
\emph{Middle panels:} Total energy (extensive) as a function of $1/L$.
\emph{Bottom panels:} Pressure versus $1/L$.
This quantity is rather insensitive to the statistics per iteration and the
basis set. 
} 
\label{s3:rs144}
\end{figure}

\begin{figure}[h]
\includegraphics[width=\columnwidth]{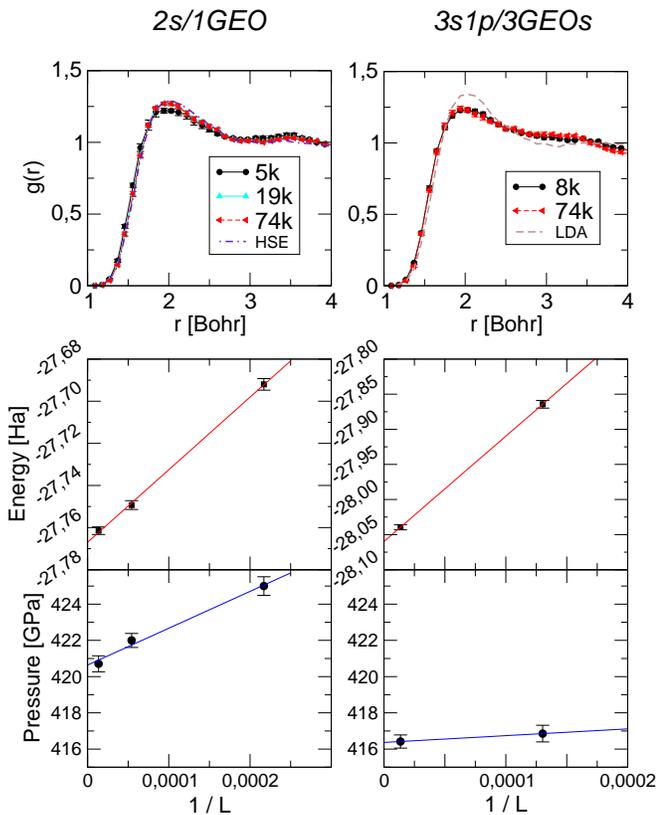}
\caption{Accuracy of the wave function: tests at density $r_s$=1.24
 and T=1700 K. The left panels refer to the $2s$/1GEO  basis set,
 while the right panels to the much more accurate
 $3s1p$/3GEOs. 
\emph{Top panels:} Radial distribution function  $g(r)$ obtained for
different $L$. 
We also plot the DFT predictions for LDA
and HSE xc functional. All the calculations are performed for 54 atoms
at the $\Gamma$ point. 
\emph{Middle panels:} Total energy (extensive) as a function of $1/L$.
The two basis extrapolate to two different
energies, the difference being of $\approx$ 5 mHa/atom. Also in this
case the extrapolation is linear.  
\emph{Bottom panels:} Pressure versus $1/L$.
This quantity is rather insensitive to $L$ and the
basis.} 
\label{s3:rs124}
\end{figure}

The results are  displayed in Figs.~\ref{s3:rs144} and \ref{s3:rs124}
for the same two densities as in Tab.~\ref{tableen}, and two
temperatures, 600 and 1700 K. 
By increasing the statistical accuracy, namely by increasing the
number $L$ of statistical samples  
used for each MD step,
the two basis considered ($2s$/1GEO and $3s1p$/3GEOs) extrapolate to
two different 
energies, the difference being more than  $\approx$ 3 mHa/atom. The
extrapolation is linear, suggesting that the bias due to the finite
statistical error is systematically improvable.  
Despite the sizable discrepancy in the $1/L\to 0$ extrapolated internal 
energies (see middle panels of Figs.~\ref{s3:rs144} and \ref{s3:rs124}), 
it  is remarkable  that the use of the small basis set   
does not lead to any significant bias in correlation functions such as pressure (accurate
within few GPa) and radial distribution functions that can be almost
superposed to the largest basis set calculation, within statistical errors. 

This test represents therefore a meaningful example where it is shown
that, by consistently optimizing  
a wave function in a small basis set, within a fully correlated
approach, the most important correlation  
functions, energy differences and derivatives, are
satisfactorily taken into account,  
opening the way for realistic  large scale calculations of correlated
materials.

\section{Conclusions}
\label{conclusions}

In this work we have introduced a novel approach for a systematic and automatic 
reduction of the basis set in wave function based approaches developed on
localized atomic orbitals. The method is built upon a
density matrix embedding scheme, constructed by partially projecting
the Slater density matrix or the AGP geminal on a given atomic site,
yielding the ``geminal embedded orbitals'' or GEOs.
The advantage of the GEO
procedure is mainly because, within our formulation, the optimal
atomic natural orbitals, which diagonalize the partially projected
density matrix, are obtained by 
solving a linear problem that
has a unique, computationally feasible (scaling as the cube of the
dimension of the basis), and automatic solution.
The embedding devised here is also shown to be superior to the
standard atomic density matrix embedding, which gives the 
``maximum-occupancy'' natural hybrid orbitals.
In the VMC framework, the method has been found very useful for the reduction 
of the number of variational parameters, which is otherwise prohibitive
for an accurate statistical optimization, as we have seen for the
water molecule and the liquid hydrogen at high pressure. The compactness of the
GEO basis set also leads to remarkable physical insights in
understanding the chemical bonds in molecules and solids, as is the
case of cerium, presented in this work. The well known volume collapse
occurring in this material is explained within this formulation as a
first-order transition of electronic character, where the GEOs
involving $f$ orbitals hybridized with $p$ states change their
localization character at the transition, being broader in the
$\gamma$ than in the $\alpha$ phase, contrary to the common
wisdom. 

We finally comment that the present technique is not only practical
to generate a more compact basis set for the Slater determinant used
in HF, DFT\cite{cp2k,siesta}  and  QMC calculations, 
but it could be extended also to other many-body approaches, by
targeting the one-body density matrix of a correlated wave function, 
rather than the geminal in the AGP form. 

Indeed, the use of a localized basis 
is well known to produce much better performing
algorithms, and the scheme we have introduced can
be easily extended to these correlated cases and further improve the efficiency
of present algorithms\cite{yambo,espresso}.  

While it is not clear
that the optimal basis set found
for the one-body density matrix is also optimal for the
one-body basis set over which the many-body wave function is
expanded, we believe that it is however a promising 
extension of the present technique.
We remark that, in the QMC framework, whenever the dynamical
correlation is treated 
by the Jastrow factor, 
the Jastrow correlated AGP ansatz recovers a substantial fraction of
correlation energy in a large variety of difficult cases, where
one does not need to go beyond the GEO scheme presented in this work.

Last but not least, the GEO basis set derived by the density matrix
embedding scheme described in this paper can be useful to generate
accurate low-energy Hamiltonians from mean-field or many-body
\emph{ab initio} calculations.

\section*{Acknowledgments}
One of us (S.S.) acknowledges support by MIUR-Cofin 2010.  
The computational ressources used for this work have
been partially provided by the PRACE project number 2012061116, and
IDRIS project number 96493. 
Part of this research has used computational resources of the
K computer provided by the RIKEN Advanced Institute for Computational
Science through the HPCI System Research Projects (hp120174 and hp140092).

\bibliography{biblio,cerium,zundel}

\end{document}